\documentclass[journal=jpcl,manuscript=extarticle,layout=twocolumn]{achemso}
\setkeys{acs}{maxauthors=0}

\makeatletter
\let\l@addto@macro\relax
\makeatother
\usepackage[fontsize=9pt]{scrextend}

\usepackage{graphicx}
\usepackage{bm}
\usepackage{xcolor}
\usepackage{siunitx}
\usepackage{physics}

\newcommand{\rd}{\mathrm{d}}

\usepackage{soul} 
\newcommand{\rev}[1]{\textcolor{black}{#1}} 
\newcommand{\onlinecite}[1]{\nocite{#1}\citenum{#1}}

\usepackage{amssymb}

\usepackage{xr-hyper}
\usepackage{hyperref}

\usepackage[symbol]{footmisc}

\usepackage{braket}
\usepackage{dcolumn} 
\newcolumntype{d}[1]{D{.}{.}{#1}} 

\title{Towards a correct description of initial electronic coherence in nonadiabatic dynamics simulations}
\author{Jonathan R.\ Mannouch}
\affiliation{Hamburg Center for Ultrafast Imaging, Universit\"at Hamburg and the Max Planck Institute for the Structure and Dynamics of Matter, Luruper Chaussee 149, 22761 Hamburg, Germany}
\email{jonathan.mannouch@mpsd.mpg.de}
\author{Aaron Kelly}
\email{aaron.kelly@mpsd.mpg.de}
\affiliation{Hamburg Center for Ultrafast Imaging, Universit\"at Hamburg and the Max Planck Institute for the Structure and Dynamics of Matter, Luruper Chaussee 149, 22761 Hamburg, Germany}

\date{\today}

\makeatletter
\newcommand*{\addFileDependency}[1]{
  \typeout{(#1)}
  \@addtofilelist{#1}
  \IfFileExists{#1}{}{\typeout{No file #1.}}
}
\makeatother

\newcommand*{\myexternaldocument}[1]{
    \externaldocument{#1}
    \addFileDependency{#1.tex}
    \addFileDependency{#1.aux}
}

\myexternaldocument{si}

\begin{document}

\maketitle

\begin{abstract}
The recent improvement in experimental capabilities for interrogating and controlling molecular systems with ultrafast coherent light sources calls for the development of theoretical approaches that can accurately and efficiently treat electronic coherence. However, the most popular and practical nonadiabatic molecular dynamics techniques, Tully's fewest-switches surface hopping and Ehrenfest mean-field dynamics, are unable to describe the dynamics proceeding from an initial electronic coherence. While such issues are not encountered with the analogous coupled-trajectory algorithms or numerically exact quantum dynamics methods, applying such methods necessarily comes with a higher computational cost. Here we show that a correct description of initial electronic coherence can indeed be achieved using methods that are based on an ensemble of independent trajectories. The key is the introduction of an initial sampling over the electronic phase space and the use of the correct observable measures, both of which are naturally achieved when working within the semiclassical mapping framework. 
\end{abstract}

\includegraphics[width=1.0\linewidth]{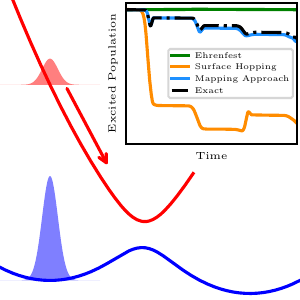}

The development and application of coherent light sources for probing and controlling the properties of matter provides substantial motivation for including the effects of the light source within computational simulations\cite{shapiro2012quantum}. For example, when simulating ultrafast laser-driven photochemical dynamics, one would like to describe the photoexcitation step on the same footing as the subsequent nonadiabatic relaxation processes. In other words, the molecular system should be initialized in the ground state and the excitation of the system should be simulated in real time through an explicit description of the pulse. It is, however, known that the most commonly used independent-trajectory approaches for simulating nonadiabatic dynamics in chemistry, Ehrenfest dynamics\cite{McLachlan1964Ehrenfest,Grunwald2009QCLE} and fewest-switches surface hopping (FSSH)\cite{Tully1990hopping,Subotnik2016review}, can fail to capture the correct light-induced coherent dynamics.\cite{Fiedlschuster2017,Mignolet2019}

As a result, most simulations indirectly take the effect of the pulse into account by initializing the system in an incoherent mixture of the photoaccessible excited electronic states\cite{Suchan2018,Barbatti2020}, with the nuclei still in their ground-state distribution. This is underpinned by two assumptions, which may or may not be valid in real photochemical scenarios. Firstly, the electromagnetic pulse is assumed to be short on the time scale of the nuclear motion, so that the nuclear wavepacket is not substantially altered by the pulse\cite{Messiah}. Secondly, decoherence is assumed to be fast, so that the resulting wavepackets on different electronic surfaces will decohere before any conical intersections are reached.\cite{Mignolet2018} In order to go beyond this commonly used computational protocol, a natural first step is to relax the second assumption and initialize simulations in the physically-relevant electronic coherence. 

Even this simple extension provides a serious challenge for the most commonly used independent-trajectory techniques.\cite{Tran2024} For example it has been recently shown that Ehrenfest and FSSH cannot describe the initial decoherence of a pair of coherent wavepackets, nor the subsequent electronic population dynamics on passing through an avoided crossing.\cite{Villaseco2024} As these methods are perhaps the most practical for treating nonadiabatic dynamics in molecular systems, there is a serious need to develop equally practical methods that can describe this situation correctly. 

One possibility is to utilize wavefunction-based approaches that calculate time-evolved observables, $\braket{\hat{B}(t)}$, according to
\begin{equation}
\label{eq:obs_wf}
\braket{\hat{B}(t)}=\int\rd\bm{q}_{t}\int\rd\bm{q}'_{t}\braket{\psi(\bm{q}_{t},t)|\hat{B}|\psi(\bm{q}'_{t},t)} ,
\end{equation}
where $\ket{\psi(\bm{q}_{t},t)}=\sum_{\lambda}c_{\lambda}(\bm{q}_{t},t)\ket{\psi_{\lambda}(\bm{q}_{t})}\ket{\bm{q}_{t}}$ is the time-evolved wavefunction, expressed in terms of an eigenstate of the nuclear position operator, $\ket{\bm{q}_{t}}$, and the adiabatic electronic states, $\ket{\psi_{\lambda}(\bm{q})}$. Such a representation encompasses approximate Gaussian wavepacket techniques\cite{Ben-Nun2000,Richings2015vMCG,Makhov2017,Curchod2018review} and coupled semiclassical trajectory approaches,\cite{donoso1998,Abedi2014MQC,Pieroni2021,Dupuy2024} derived for example via the exact factorization.\cite{Abedi2010} A recent paper\cite{Villaseco2024} has shown that coupled trajectories can indeed alleviate the problems associated with Ehrenfest and FSSH when starting in an initial electronic coherence, albeit at a higher computational cost. 

Here we assess the applicability of the semiclassical mapping formalism\cite{Miller1978mapping,Stock1997mapping,Stock2005nonadiabatic,Kim2008Liouville,Kelly2012mapping,spinmap} to this problem. The mapping formalism  provides a way of going beyond the standard approaches of Ehrenfest and FSSH, without the need to invoke coupled trajectory simulation algorithms. Within this formalism, independent-trajectory approaches can be derived by making approximations to real-time correlation functions of the form
\begin{equation}
\label{eq:corr}
\braket{\hat{B}(t)}=\int\rd\bm{q}\int\rd\bm{p}\,\tr[\hat{\rho}^{\mathrm{W}}(\bm{q},\bm{p})\hat{B}^{\mathrm{W}}(\bm{q},\bm{p},t)] ,
\end{equation}
where $\hat{B}^{\mathrm{W}}(\bm{q},\bm{p})$ is the partial Wigner transform\cite{Kapral1999MQCD} of the operator $\hat{B}$ with respect to the nuclear degrees of freedom and $\tr[\cdots]$ denotes a quantum trace over the electronic degrees of freedom. Additionally the initial state of the system is expressed in terms of the density matrix, $\hat{\rho}=\int\rd\bm{q}\,\rd\bm{q}'\ket{\psi(\bm{q},0)}\bra{\psi(\bm{q}',0)}$. This provides an additional framework through which nonadiabatic dynamics and decoherence phenomena can be understood. 

\begin{figure*}
\centering
\includegraphics[width=1.0\linewidth]{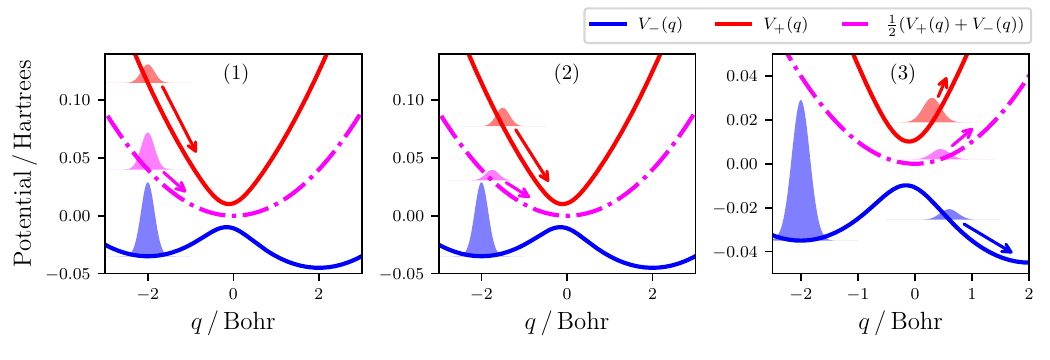}
    \caption{A schematic illustrating certain aspects of the dynamics generated from an initially coherent wavepacket between the ground (blue) and excited (red) Born-Oppenheimer surfaces. The different components of the initial electronic reduced density matrix are represented by the Gaussians in panel (1), with the arrows showing their instantaneous motion. The red and blue Gaussians correspond to the associated adiabatic populations and the pink Gaussian to the electronic coherence. Panel (2) illustrates the decoherence of the initially coherent wavepacket and panel (3) illustrates a nonadiabatic transition, occurring when the excited-state wavepacket passes through the coupling region. The Born-Oppenheimer surfaces in each panel correspond to the same model as considered in Ref.~\protect\onlinecite{Pieroni2021,Villaseco2024}.}    
    \label{fig:model}
\end{figure*}
In order to better understand the difficulty in describing the dynamics of an initial electronic coherence, it is instructive to first consider what the correct dynamics should look like. In panel (1) of Fig.~\ref{fig:model} we illustrate a typical photochemical scenario, where an ultra-short laser pulse has promoted a small fraction of a stable ground-state wavepacket (blue Gaussian) to the excited Born-Oppenheimer electronic surface (red Gaussian). Crucially, because the Born-Oppenheimer surfaces are typically far apart at the Franck-Condon geometry in photochemical systems, the initial dynamics can be decomposed into independent motions associated with each component of the electronic reduced density matrix \footnote{In the following, we refer to the diagonal elements of the density matrix as populations and the off-diagonal elements as coherences.}. In the classical-nuclear limit, the electronic populations (corresponding to the blue and red Gaussians in Fig.~\ref{fig:model}) evolve on their associated Born-Oppenheimer surface, while the electronic coherences (corresponding to the pink Gaussian in Fig.~\ref{fig:model}) evolve on the average of the surfaces\cite{Egorov1999vibronic,Shi2004goldenrule,Shi2005nonadiabatic} (given by the dashed pink line in Fig.~\ref{fig:model}). 

Applying this to the scenario illustrated in Fig.~\ref{fig:model}, we see that the ground-state population will remain stationary in its potential well, while the excited-state population and the coherences will initially experience a force towards the right. As the excited- and ground-state wavepackets separate, the two will decohere, resulting in a decay of the initial electronic coherence (panel (2) of Fig.~\ref{fig:model}). Later, when the excited-state wavepacket reaches the coupling region at $q=0$, a nonadiabatic transition will then promote part of this wavepacket onto the ground-state surface, momentarily recreating electronic coherence (panel (3) of Fig.~\ref{fig:model}). Further nonadiabatic transitions occur whenever wavepackets recross the coupling region, which can also lead to `recoherence' events as wavepackets on different surfaces overlap.

In order to describe all aspects of these dynamics with independent trajectories, the following two minimal criteria must be satisfied. Firstly, in order to describe the initial independent motion of the red and blue wavepackets, trajectories must initially feel the force of either the ground- or excited-state surface, with a ratio that matches the associated electronic populations. Secondly, in order to describe the initial dynamics of the electronic coherence, some trajectories must also propagate on the average surface.   

The problem with Ehrenfest and FSSH is that they do not simultaneously fulfill both of these criteria. While the mean-field force used in Ehrenfest dynamics is suitable for describing the dynamics of the electronic coherence, it fails to correctly propagate the electronic populations on single Born-Oppenheimer surfaces. In contrast, while the surface-hopping force of FSSH guarantees that the electronic populations are correctly propagated on single Born-Oppenheimer surfaces, the electronic coherences are no longer propagated on the average surface. FSSH additionally suffers from a so-called `inconsistency error,'\cite{Subotnik2016review} which arises from the fact that the active propagation surface can become inconsistent with the underlying electronic wavefunction during the dynamics. 

\begin{figure*}[t]
   \centering
    \includegraphics[width=1.0\linewidth]{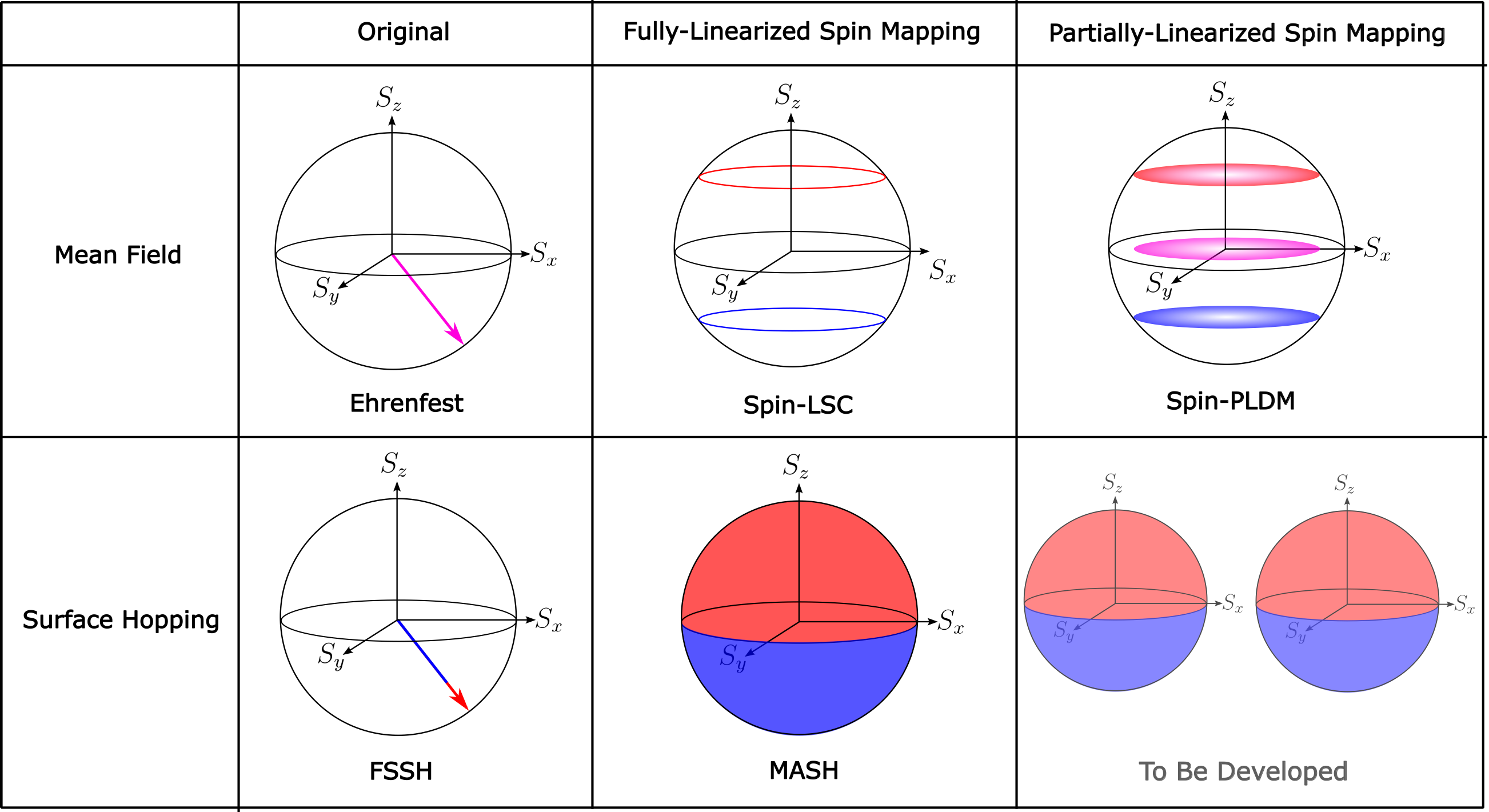}
    \caption{A schematic illustrating the phase-space regions from which the electronic spin-mapping variables are initialized for various independent-trajectory approaches. The red and blue shading signifies regions for which the trajectories would initially feel a force corresponding to the upper and lower adiabatic surfaces respectively and the pink shading signifies an initial force that is an average of the two surfaces. For spin-PLDM, the initial distribution of the average of the two sets of spin-mapping variables is given. See the supplementary material for more details.}
    \label{fig:spin}
\end{figure*}

One way to resolve these issues is to utilize semiclassical mapping approaches. While several different mappings have been suggested,\cite{Meyer1979spinmatrix,Stock1997mapping,Liu2019} in this paper we focus on methods derived within the spin-mapping formalism.\cite{spinmap,multispin} This maps a two-state electronic subsystem onto a spin-$\tfrac{1}{2}$ particle and represents any two-state electronic wavefunction by a spin vector, $\bm{S}$, on the three-dimensional Bloch sphere. For example, the north and south poles of the Bloch sphere ($S_{z}=\pm1$ and $S_{x}=S_{y}=0$) correspond to the excited- and ground-state adiabats and the spin vector corresponding to the initial coherent wavepacket shown in panel (1) of Fig.~\ref{fig:model} is given by the left column of Fig.~\ref{fig:spin}.

As with other mapping approaches, the spin-mapping framework has enabled the development of more accurate independent-trajectory approaches that go beyond Ehrenfest and FSSH\cite{spinmap,spinPLDM1,Mannouch2023MASH}. We briefly introduce these approaches here, although more details are given in the supplementary information (SI). Spin mapping was first used to develop a more accurate mean-field approach, called spin-LSC\cite{spinmap}, which utilizes a larger spin sphere of $\sqrt{3}\bm{S}$ in order to reproduce the correct spin magnitude of a quantum spin-$\tfrac{1}{2}$ particle. On this larger sphere, the spin vectors that represent the two adiabatic states (i.e. $\sqrt{3}S_{z}=\pm1$) lie on the two polar circles, as illustrated in the upper-middle panel of Fig.~\ref{fig:spin}. These polar circles constitute the initial sampling regions for the spin-LSC approach. By sampling from the polar circles with the correct weighting, each spin-LSC trajectory has an initial force that corresponds to a single Born-Oppenheimer surface, introducing an essential feature lacking in Ehrenfest dynamics. An initial electronic coherence is also simultaneously describable, because any point on the polar circles generally has a non-zero value for $S_{x}$ and $S_{y}$.  

Most recently, the spin-mapping framework was used to develop a more accurate surface-hopping approach, called the mapping approach to surface hopping (MASH)\cite{Mannouch2023MASH,MSMASH,Runeson2023MASH}. The main difference between FSSH and MASH is how the active surface for the propagation  of the nuclei is determined. While the active surface in FSSH is switched stochastically according to the time evolution of the underlying electronic wavefunction, in MASH the active surface is instead chosen according to the spin-hemisphere in which the spin vector currently resides, as illustrated in the lower-middle panel of Fig.~\ref{fig:spin}. Physically, this corresponds to setting the active surface as the adiabat for which the electronic wavefunction has the highest associated probability. As a result, MASH has a purely deterministic dynamics, where the active surface changes whenever the spin vector crosses the equator. This guarantees that the MASH active surface is always consistent with the electronic wavefunction, thereby avoiding the inconsistency error of FSSH that is known to significantly degrade its accuracy.\cite{Subotnik2016review} 

So far we have only considered fully-linearized mapping-based approaches,\cite{Sun1998mapping,Miller2001SCIVR,Shi2004goldenrule} which contain a single set of electronic mapping variables, and are generally able to describe the dynamics associated with the electronic populations. In order to correctly describe the dynamics of the coherences, partially-linearized approaches are generally needed.\cite{Miller2012coherence,nonlinear} These approaches use two sets of mapping variables, with each describing the electronic dynamics generated by either the forward or backward propagator.\cite{Hsieh2012FBTS,Huo2012PLDM,spinPLDM1} The partially-linearized version of spin-LSC is called spin-PLDM.\cite{spinPLDM1,spinPLDM2} Each set of spin-mapping variables in spin-PLDM is sampled independently from the same polar circles as in spin-LSC, such that the average of the two sets is distributed according to the upper-right panel of Fig.~\ref{fig:spin}. In particular, spin-PLDM trajectories can now be initialized with a force corresponding to the average of the two Born-Oppenheimer surfaces (pink region), which occurs whenever the two sets of spin-mapping variables are initialized on different polar circles. While a partially-linearized version of MASH could be formulated in principle, this has yet to be developed. 

In order to assess the ability of the different semiclassical mapping approaches to describe the nonadiabatic dynamics of an initial electronic coherence, we consider a typical photochemical scenario using the one-dimensional model system employed in Refs.~\onlinecite{Pieroni2021,Villaseco2024}. The Born-Oppenheimer surfaces and the initial coherent electronic wavepacket are depicted in panel (1) of Fig.~\ref{fig:model}. The initial state has 80\% of its weight on the lower adiabat and its Gaussian profile corresponds to the ground vibrational eigenstate of the potential well located at $q\approx-2$, all within the harmonic approximation. More details about the model are given in the SI. 

\begin{figure*}[t]
   \centering
    \includegraphics[width=1.0\linewidth]{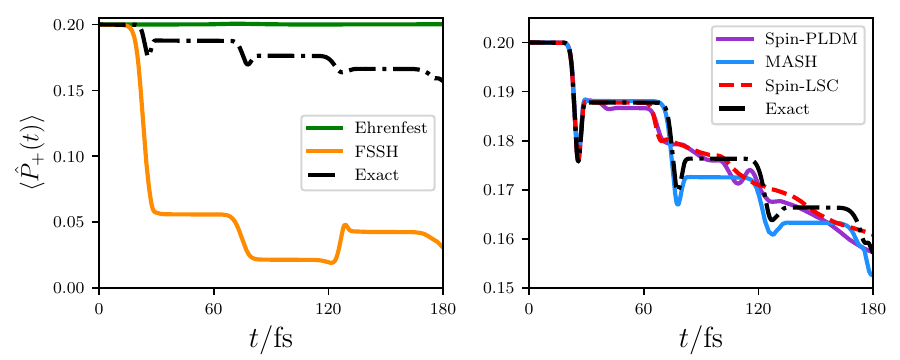}
    \caption{The population of the upper adiabatic electronic state as a function of time, calculated with a range of different independent-trajectory techniques. Our Ehrenfest and FSSH results match those presented in Ref.~\protect\onlinecite{Villaseco2024}.}
    \label{fig:pop}
\end{figure*}

We first consider the time evolution of the excited electronic-state population, which can be expressed as a single real-time correlation function of the form of Eq.~\ref{eq:corr} with $\hat{B}=\hat{P}_{+}(\bm{q})$. This means that the semiclassical mapping approaches can be used to calculate this quantity directly, with the results given in Fig.~\ref{fig:pop}. From the potentials shown in Fig.~\ref{fig:model}, the wavepacket in the excited adiabatic state should oscillate about the coupling region at $q=0$. As the parameters for this model are close to the Born-Oppenheimer limit, only a small amount of population is transferred at each crossing, giving rise to the step-like behaviour of the exact population dynamics in Fig.~\ref{fig:pop}, which were computed using a split-operator approach.\cite{Tannor}

In agreement with previous work,\cite{Villaseco2024} the left panel of Fig.~\ref{fig:pop} shows that both Ehrenfest and FSSH are unable to describe the population dynamics originating from such an initial coherent electronic state. The composition of the initial state means that the major contribution to the Ehrenfest mean-field force comes from the ground-state surface, so that almost none of the Ehrenfest trajectories reach the coupling region. This explains why Ehrenfest does not give rise to any population transfer in this case. While FSSH does propagate the right fraction of trajectories on the excited-state surface, many of the trajectories nevertheless have an inconsistent electronic wavefunction (again due to the composition of the initial coherent state) which leads to the large error seen in the populations.  

In contrast, all of the spin-mapping approaches closely match the exact result. This improvement stems from the initial sampling introduced over the electronic spin-mapping variables. For the mean-field methods of spin-LSC and spin-PLDM, the initial sampling over the polar circles [Fig.~\ref{fig:spin}] leads to the correct fraction of trajectories experiencing the excited-state force and therefore reaching the coupling region. For MASH, the trajectories that propagate on the upper surface are those that are initialized with spin-mapping variables in the upper hemisphere, so that the electronic wavefunction is consistent with the nuclear propagation surface. Of all the spin-mapping methods, MASH most accurately reproduces the step-like behaviour of the exact populations at longer times. This is because the MASH force always comes from single Born-Oppenheimer surfaces, in contrast to the spin-mapping mean-field methods, where this is only initially the case.     

We now consider the time-evolution of the electronic coherences. One simple measure of the magnitude of the electronic coherence between two adiabatic states is obtained from the associated off-diagonal element of the electronic reduced density matrix, $\rho_{-+}(t)$. Unlike the population observable considered above, $|\rho_{-+}(t)|$ is not expressible as a single real-time correlation function. Therefore in order to calculate this quantity with semiclassical mapping approaches, it must first be re-expressed in terms of correlation functions of the form of Eq.~\ref{eq:corr}, which for a two-state system can be achieved as follows
\begin{equation}
\label{eq:coh1_corr}
|\rho_{-+}(t)|=\frac{1}{2}\sqrt{\braket{\hat{\sigma}_{x}(\bm{q},t)}^{2}+\braket{\hat{\sigma}_{y}(\bm{q},t)}^{2}} .
\end{equation}
Here, $\hat{\bm{\sigma}}(\bm{q})$ are the Pauli spin matrices expressed in the adiabatic basis. Hence any ensemble of independent trajectories should first be used to compute the correlation functions $\braket{\hat{\sigma}_{x}(\bm{q},t)}$ and $\braket{\hat{\sigma}_{y}(\bm{q},t)}$, and these quantities can subsequently be inserted into Eq.~\ref{eq:coh1_corr} to obtain the coherence measure.

\begin{figure*}[t]
   \centering
    \includegraphics[width=1.0\linewidth]{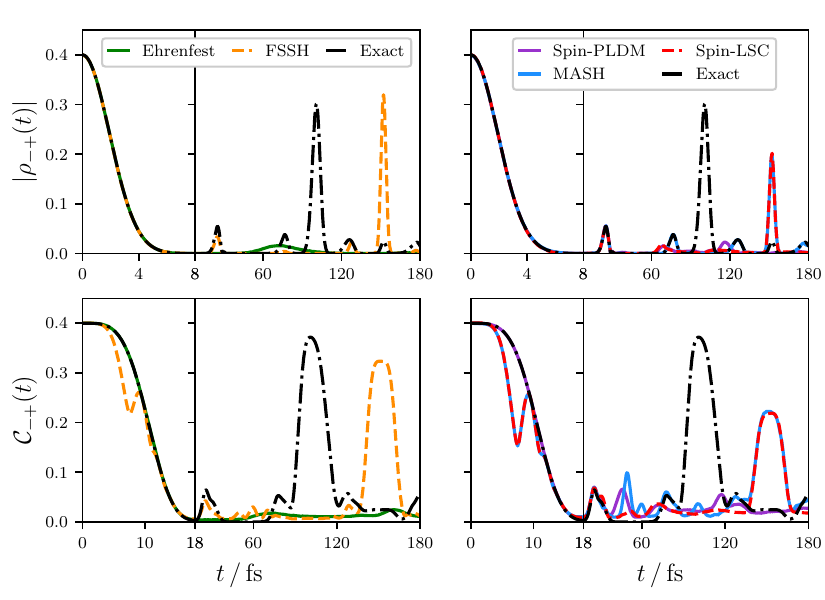}
    \caption{Two different electronic coherence measures, $|\mathcal{\rho}_{-+}(t)|$ and $\mathcal{C}_{-+}(t)$, calculated with a range of different independent-trajectory techniques. The mathematical expressions for these measures are given by Eqs.~(\ref{eq:coh1_corr}) and (\ref{eq:coh2}), respectively. Our $\mathcal{C}_{-+}(t)$ measure is strongly related to that used in Refs.~\protect\onlinecite{Vacher2017,Curchod2018,Villaseco2024,Villaseco2024_2}.}
    \label{fig:coh}
\end{figure*}

The upper panels of Fig.~\ref{fig:coh} give this coherence measure computed for the same model previously considered for the electronic populations. Starting with the initial decoherence process, the exact dynamics show that the initial electronic coherence decays as the excited-state wavepacket moves apart from the wavepacket on the ground state [panel (2) of Fig.~\ref{fig:model}].
The upper panels in Fig.~\ref{fig:coh} show that all of the independent-trajectory approaches are able to describe this initial decoherence behaviour correctly. This may seem surprising given that individual trajectories are known to remain `overcoherent' after a nonadiabatic transition.\cite{Bittner1995Decoherence,Fang1999FSSH_inconsistency,Subotnik2011decoherence} Indeed, if Eq.~\ref{eq:coh1_corr} was calculated using just a single trajectory, $|\rho_{-+}(t)|$ would be constant as a function of time and no decoherence would be observed. Decoherence therefore arises as a result of ensemble averaging. While the contribution to $\braket{\hat{\sigma}_{x}(\bm{q},t)}$ and $\braket{\hat{\sigma}_{y}(\bm{q},t)}$ from each trajectory is in general non-zero, the sign can vary, leading to phase cancellation\cite{Fiete2003} among trajectories such that the ensemble-averaged correlation functions can be zero. This highlights an important philosophical point underpinning most trajectory-based approaches: physical meaning can only be ascribed to average quantities derived from the trajectory ensemble, and not to the individual trajectories themselves. While individual trajectories do remain unphysically overcoherent after a nonadiabatic transition, Fig.~\ref{fig:coh} highlights that the trajectory ensemble displays the correct decoherence behaviour for a wide range of methods. This is true also for the FSSH results, which were obtained without needing to apply decoherence corrections.\footnote{However we would also like to clarify that this does not mean that decoherence corrections are never needed in FSSH. There are well known cases where  decoherence corrections are required to even ensure the trajectory ensemble is physically correct.\cite{MASHrates}}

While $|\rho_{-+}(t)|$ is useful due to its close connection with linear spectroscopic signatures,\cite{nonlinear} it nevertheless does not offer the most rigorous definition of electronic coherence. In order to see this, $|\rho_{-+}(t)|$ can be expressed in terms of the time-dependent wavefunction coefficients, $c_{j}(\bm{q}_{t},t)$, as
\begin{equation}
\label{eq:coh1_wf}
|\rho_{-+}(t)|=\left|\int\rd\bm{q}_{t}\,c^{*}_{+}(\bm{q}_{t},t)c_{-}(\bm{q}_{t},t)\right| ,
\end{equation}
which are themselves defined by the expression for the wavefunction appearing in Eq.~\ref{eq:obs_wf}. Because the coefficients are not positive functions of $\bm{q}$, $|\rho_{-+}(t)|$ can in principle be zero even if the coefficients on different surfaces are spatially overlapping. To resolve this issue, the order of the integral and the magnitude in Eq.~\ref{eq:coh1_wf} can be interchanged to give\cite{Vacher2017,Curchod2018,Villaseco2024,Villaseco2024_2}
\begin{equation}
\label{eq:coh2}
\begin{split}
\mathcal{C}_{-+}(t)&=\int\rd\bm{q}_{t}\,\Big|c^{*}_{+}(\bm{q}_{t},t)c_{-}(\bm{q}_{t},t)\Big| \\
&=\frac{1}{2}\int\rd\bm{q}_{t}\sqrt{\Braket{\hat{\sigma}^{(\bm{q}_{t})}_{x}(t)}^{2}+\Braket{\hat{\sigma}^{(\bm{q}_{t})}_{y}(t)}^{2}} ,
\end{split}
\end{equation}
where $\hat{\sigma}^{(\bm{q})}_{j}=\hat{\sigma}_{j}(\bm{q})\ket{\bm{q}}\!\bra{\bm{q}}$ is a product of a Pauli spin matrix and a projector onto the specific eigenstate of the nuclear position operator, $\ket{\bm{q}}$. For the independent-trajectory approaches, the coherence correlation functions entering into Eq.~\ref{eq:coh2} can be calculated by histogramming the appropriate  spin-mapping variable with respect to the nuclear position, $\bm{q}(t)$ (see the SI for more details).

We proceed by considering the initial dynamics of this coherence measure, shown in the lower panels of Fig.~\ref{fig:coh}. Now not all of the independent-trajectory approaches are able to exactly describe the decay of this measure. The approaches that can (i.e., Ehrenfest and spin-PLDM) are those that initially propagate at least some of their trajectories on the average surface. This highlights one of the major advantages of partially-linearized approaches, like spin-PLDM, which are able to simultaneously describe the population and coherence dynamics relatively accurately by initially propagating trajectories on both single and average Born-Oppenheimer surfaces. 

For all of the independent-trajectory approaches, the ability to describe the longer-time coherence dynamics for both measures is mixed. We first consider the smaller transient coherence peaks, which originate in the exact dynamics from nonadiabatic transitions (see panel (3) of Fig.~\ref{fig:model}). The first such peak is well captured by all of the mapping-based approaches and subsequent peaks are almost perfectly captured by MASH. As expected, this finding mirrors the relative performance of the methods at capturing the correct population dynamics [Fig.~\ref{fig:pop}].  

Notably, none of the methods are able to reproduce the large recoherence peak at $\approx 100$ fs, which arises from the re-overlap of the excited-state wavepacket with the stationary ground-state wavepacket at $q\approx -2$. Additionally, most of the independent-trajectory approaches show several spurious recoherence peaks in the $\mathcal{C}_{-+}(t)$ measure, which have no analogue in the exact dynamics.

\begin{figure*}[t]
   \centering
    \includegraphics[width=1.0\linewidth]{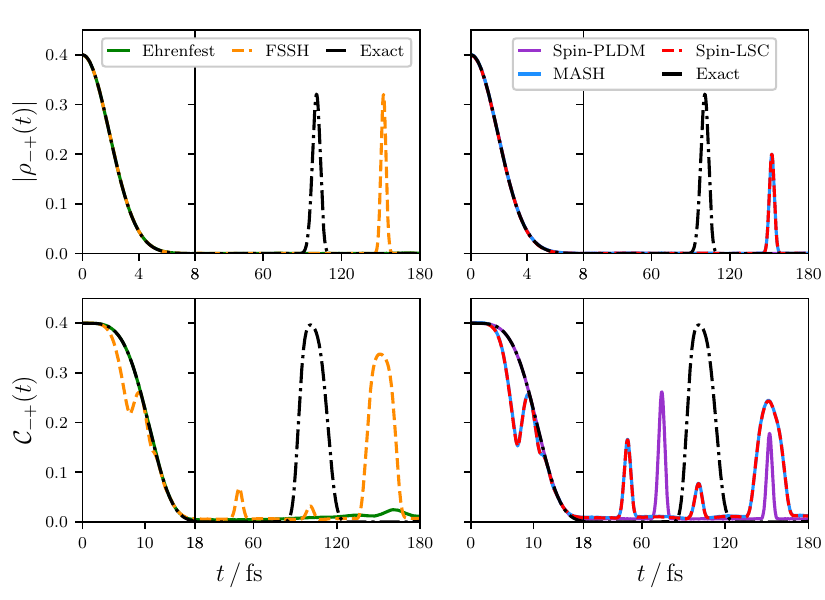}
    \caption{The Born-Oppenheimer limit of the coherence measures, $|\mathcal{\rho}_{-+}(t)|$ and $\mathcal{C}_{-+}(t)$, shown in Fig.~\ref{fig:coh}.}
    \label{fig:coh_BO}
\end{figure*} 
In order to better understand the source of these discrepancies, we also consider the coherence dynamics in the Born-Oppenheimer limit, with the results given in Fig.~\ref{fig:coh_BO}. Note that the recoherence behaviour of all of the independent-trajectory methods in the Born-Oppenheimer limit largely matches their behaviour in the full system. In particular, spin-PLDM reduces to the so-called Wigner-averaged classical limit (WACL)\cite{Egorov1999vibronic,Shi2004goldenrule,Shi2005nonadiabatic} in the Born-Oppenheimer limit.\cite{nonlinear} Given that WACL only differs from the exact dynamics as a result of its classical-nuclear approximation, this suggests that the classical-nuclear approximation is the main reason behind why the recoherences are difficult to describe with independent-trajectories.  

In conclusion, we have assessed the ability of a range of independent-trajectory nonadiabatic dynamics approaches to describe the dynamics proceeding from a coherent electronic state using a model relevant for photochemical systems. The ability to treat these initial conditions provides a first step towards a more accurate description of the effects of tailored electromagnetic pulses within nonadiabatic simulations, which is a crucial step for providing a more direct connection with cutting-edge experiments.   

In order to correctly describe an initial electronic coherence with independent-trajectory approaches, we have found that it is important to introduce an initial sampling over the electronic phase space, as is naturally incorporated within the semiclassical mapping framework. In particular we considered all possible flavours of the spin-mapping approach, all of which were found to lead to the right fraction of trajectories being initialised on each adiabatic surface, as required to reproduce the correct population dynamics. MASH was particularly successful for this model, where the surface-hopping force meant that the approach could accurately describe the nonadiabatic transitions associated with multiple crossings of the localised coupling region. However, all of the spin-mapping methods were seen to be a significant upgrade on the more commonly used Ehrenfest and FSSH approaches, which in contrast failed to correctly reproduce these aspects of the dynamics. While we have exclusively focused on the spin-mapping formalism here, our conclusions would have been identical for the majority of other mapping representations.

We also showed that independent-trajectory approaches can reproduce the correct decoherence behaviour. Even though the individual trajectories of these approaches remain overcoherent, decoherence is nevertheless included via phase-cancellation between them through ensemble averaging in the construction of the relevant correlation functions. However, not every independent-trajectory approach was able to perfectly describe all aspects of the coherence measures. In particular, for the more stringent $\mathcal{C}_{-+}(t)$ measure, it was necessary to propagate at least some trajectories on the average Born-Oppenheimer surface in order to completely capture the initial decay. This illustrates one of the advantages of partially-linearized approaches over their fully-linearized counterparts.\cite{Miller2012coherence} Similar findings were also found in previous work when calculating real-time dipole-dipole correlation functions for optical spectra, where partially-linearized approaches also offered a significant advantage.\cite{nonlinear} Given the advantages of describing the electronic population dynamics in photochemical simulations with MASH over analogous mean-field approaches, it is expected that similar advantages would also arise from a partially-linearized version of MASH for computing optical spectra in such systems. Obtaining such a method is one aspect of future work.  

Another aspect of future work is to develop mapping-based approaches that are able to couple an explicit electromagnetic pulse. For mean-field approaches this extension is straightforward and has already been achieved for fully and partially linearized mapping approaches such as PBME and FBTS,\cite{Hanna_2013} as well as spin-LSC.\cite{ultrafast} We are currently working on the necessary developments of the algorithm to include this effect correctly for surface-hopping methods like MASH. 

Finally, the classical-nuclear approximation is observed to lead to the absence of recoherence phenomena in the dynamics of independent-trajectory approaches. In the majority of realistic (high-dimensional) systems, recoherence phenomena are suppressed, and one may expect that this deficiency will not be a major problem. However, the classical-nuclear approximation is also known to `wash out' other coherence related phenomena, such as contributions to dipole-dipole correlation functions that give rise to vibronic progressions in optical spectra.\cite{McRobbie2009nonlinear,Karsten2018vibronic,lively2021simulating} For this application, it would therefore be useful to develop extensions of nonequilibrium trajectory-based approaches that go beyond the classical-nuclear approximation. While a number of ring-polymer extensions to nonadiabatic trajectory-based approaches have been developed for simulating equilibrium dynamics,\cite{Shushkov2012RPSH,mapping,Ananth2013MVRPMD,vibronic,Chowdhury2017CSRPMD,Bossion2021NRPMD,Bossion2023NRPMD} to our knowledge there are currently no established methods to tackle the nonequilibrium regime.    

\rev{
\textbf{Supporting Information:}
The supplementary material provides all of the necessary information to reproduce the results in the main text.
}

\begin{acknowledgement}
This work was supported by the Cluster of Excellence ``CUI: Advanced Imaging of Matter'' of the Deutsche Forschungsgemeinschaft (DFG) – EXC 2056 – project ID 390715994.
\end{acknowledgement}

\bibliography{references,extra_refs}

\end{document}


\maketitle


\renewcommand{\thepage}{S\arabic{page}}
\renewcommand{\theequation}{S\arabic{equation}}
\renewcommand{\thefigure}{S\arabic{figure}}
\renewcommand{\thetable}{S\arabic{table}}
\renewcommand{\thesection}{S\arabic{section}}
\renewcommand{\thesubsection}{S\arabic{section}.\arabic{subsection}}

\subsection{The Model}
We consider the same one-dimensional model used in Ref.~\onlinecite{Pieroni2021,Villaseco2024}. In the diabatic basis, the Hamiltonian is given in atomic units by
\begin{equation}
\hat{H}=\frac{p^{2}}{2m}+\bar{V}(q)\hat{\mathcal{I}}+\Delta(q)\hat{\sigma}^{\mathrm{diab}}_{x}+\kappa(q)\hat{\sigma}^{\mathrm{diab}}_{z} ,
\end{equation}
 where $q$ and $p$ are the nuclear position and momentum, $m=20000$ is the nuclear mass and $\hat{\bm{\sigma}}^{\mathrm{diab}}$ are the standard Pauli spin matrices expressed in the diabatic basis, which along with the identity operator, $\hat{\mathcal{I}}$, form a complete set of Hermitian operators for two-state systems. The diabatic Hamiltonian parameters are
\begin{subequations}
\begin{align}
\bar{V}(q)&=\tfrac{1}{2}kq^{2} , \\
\Delta(q)&=b\eu{-a\left(q-\frac{\epsilon}{2k\bar{q}}\right)^{2}} , \\
\kappa(q)&= k\bar{q}q-\tfrac{1}{2}\epsilon ,
\end{align}
\end{subequations}
where $k=0.02$, $b=0.01$, $a=3$, $\bar{q}=-2$ and $\epsilon=0.01$. The quantities used here are related to those used in Ref.~\onlinecite{Pieroni2021,Villaseco2024} by: $q=R-\tfrac{1}{2}(R_{1}+R_{2})$ and $\bar{q}=\tfrac{1}{2}(R_{2}-R_{1})$. From the diabatic Hamiltonian parameters, the adiabatic potential energies, $V_{\pm}(q)$, and the nonadiabatic coupling vector (NACV), $d(q)$, can be easily calculated using
\begin{subequations}
\begin{align}
V_{z}(q)&=\sqrt{\Delta(q)^{2}+\kappa(q)^{2}} , \\
d(q)&=\frac{\Delta(q)\frac{\partial\kappa(q)}{\partial q}-\kappa(q)\frac{\partial\Delta(q)}{\partial q}}{2(\Delta(q)^{2}+\kappa(q)^{2})} ,
\end{align}
\end{subequations}
where $V_{\pm}=\bar{V}(q)\pm V_{z}(q)$ and the $+$ and $-$ indices refer to the upper and lower states.

As in Ref.~\onlinecite{Villaseco2024}, the dynamics is initialized in a product state corresponding to the electronic wavefunction, $\ket{\psi(q)}=\sqrt{\tfrac{1}{2}(1-\Delta P)}\ket{\psi_{+}(q)}+\sqrt{\tfrac{1}{2}(1+\Delta P)}\ket{\psi_{-}(q)}$, and the nuclear ground state of the diabatic potential, $\bar{V}(q)-\kappa(q)$. The partial Wigner transform of the initial density matrix is then
\begin{subequations}
\label{eq:init}
\begin{align}
\hat{\rho}^{\mathrm{W}}(q,p)&\approx\hat{\rho}_{\mathrm{el}}(q)\rho_{\mathrm{nuc}}(q,p) , \\
\hat{\rho}_{\mathrm{el}}(q)&=\ket{\psi(q)}\bra{\psi(q)}=\tfrac{1}{2}\left[\hat{\mathcal{I}}-\Delta P\hat{\sigma}_{z}(q)+\sqrt{1-\Delta P^{2}}\,\hat{\sigma}_{x}(q)\right] , \\
\rho_{\mathrm{nuc}}(q,p)&=\frac{1}{\pi}\eu{-\frac{k(q-\bar{q})^{2}}{\omega}}\eu{-\frac{p^{2}}{m\omega}} , \label{eq:nuc_dens}
\end{align}
\end{subequations}
where $\Delta P=0.6$ is the electronic population difference between the lower and upper adiabatic states and $\omega=\sqrt{k/m}$ is the harmonic frequency. Additionally, $\hat{\bm{\sigma}}(q)$ are the Pauli spin 
matrices defined in the adiabatic basis, which are related to $\hat{\bm{\sigma}}^{\mathrm{diab}}$ by a unitary transformation.\cite{Mannouch2023MASH}

\section{Quasiclassical Trajectory Techniques}
In this section, we define the equations of motion and observables associated with all of the quasiclassical-trajectory techniques used in the main paper. The electronic spin-mapping variables, which specify the electronic state of the system, are related to the electronic wavefunction coefficents for two-state systems as follows
\begin{subequations}
\begin{align}
S_{x}&=2\mathrm{Re}[c_{+}^{*}c_{-}] , \\
S_{y}&=2\mathrm{Im}[c_{+}^{*}c_{-}] , \\
S_{z}&=|c_{+}|^{2}-|c_{-}|^{2}.
\end{align}
\end{subequations}
The spin-mapping phase space that corresponds to correctly normalized electronic wavefunctions is the surface of the Bloch sphere, which satisfies $|\bm{S}|=\sqrt{S_{x}^{2}+S_{y}^{2}+S_{z}^{2}}=1$. We therefore specify the following phase-space averages over the electronic and nuclear variables
\begin{subequations}
\begin{align}
\Braket{\cdots}_{\mathrm{FL}}&=\int\rd q\int\rd p\int\rd\bm{S}\,\delta\!\left(|\bm{S}|-1\right)\,\cdots , \\
\Braket{\cdots}_{\mathrm{PL}}&=\int\rd q\int\rd p\int\rd\bm{S}^{(\mathrm{f})}\int\rd\bm{S}^{(\mathrm{b})}\,\delta\!\left(|\bm{S}^{(\mathrm{f})}|-1\right)\delta\!\left(|\bm{S}^{(\mathrm{b})}|-1\right)\,\cdots ,
\end{align}
\end{subequations}
which are appropriate for both fully-linearized (FL) and partially-linearized (PL) approaches, containing one and two sets of spin-mapping variables respectively. Additionally, $\delta(x)$ is the Dirac delta function.

The dynamical phase-space variables are evolved for each trajectory according to the following equations of motion
\begin{subequations}
\label{eq:dynamics}
\begin{align}
\dot{S}_{x}&=\frac{2d(q)p}{m}S_{z}-2V_{z}(q)S_{y} , \label{eq:dynam_Sx} \\
\dot{S}_{y}&=2V_{z}(q)S_{x} , \label{eq:dynam_Sy} \\
\dot{S}_{z}&=-\frac{2d(q)p}{m}S_{x} , \label{eq:dynam_Sz} \\
\dot{q}&=\frac{p}{m} , \\
\dot{p}&=-\mathcal{F}(q,\bm{S}) ,
\end{align}
\end{subequations}
where for the partially-linearized approaches, $\bm{S}=\tfrac{1}{2}(\bm{S}^{(\mathrm{f})}+\bm{S}^{(\mathrm{b})})$ is the centroid (i.e., the average of the two sets of variables).

The expression for the nuclear force, $\mathcal{F}(q,\bm{S})$, the initial sampling of the electronic spin-mapping variables and the representation of the observable operators differ between the various quasiclassical approaches for nonadiabatic dynamics simulations. We consider these aspects of the various approaches in the following.
\subsection{Surface-Hopping Approaches}
Surface-hopping approaches propagate the nuclei on a single adiabatic surface at any given time. The propagation surface, $n_{\mathrm{act}}$, is referred to as the `active surface' and for the two-state system considered here, we refer to $n_{\mathrm{act}}=1$ as the upper surface and $n_{\mathrm{act}}=-1$ as the lower. The nuclear force for surface-hopping approaches is therefore given by
\begin{equation}
	\mathcal{F}^{\mathrm{SH}}(q,\bm{S})=-\frac{\partial\bar{V}(q)}{\partial q}-\frac{\partial V_{z}(q)}{\partial q}n_{\mathrm{act}} .
\end{equation}
In order to describe nonadiabatic transitions, the active surface must be dynamically altered in a way that mimics the dynamics of the underlying electronic wavefunction. How this is achieved depends on the specifics of the method. 

The most accurate way of measuring electronic observables with a surface-hopping approach is to determine the electronic coherences with the electronic wavefunction and the adiabatic populations using the active surface
\begin{subequations}
\begin{align}
		\sigma^{\mathrm{SH}}_{j}(\bm{S},n_{\mathrm{act}})&=\begin{cases} S_{j}\quad&\mathrm{for}\quad j\neq z, \\
		n_{\mathrm{act}}&\mathrm{for}\quad j=z , \end{cases} \\
        \mathcal{I}^{\mathrm{SH}}(\bm{S},n_{\mathrm{act}})&=1 ,
\end{align}
\end{subequations}
which corresponds to the density matrix approach for calculating dynamical observables.\cite{Landry2013FSSH,Subotnik2016review}

\subsubsection{Fewest-Switches Surface Hopping (FSSH)}
Fewest-switches surface hopping (FSSH)\cite{Tully1990hopping,Subotnik2016review} approximates the correlation function associated with the initial density matrix $\hat{\rho}$ [Eq.~\ref{eq:init}] and the observable $\hat{B}$ at time $t$ as
\begin{equation}
	\Tr[\hat{\rho}\eu{i\hat{H}t}\hat{B}\eu{-i\hat{H}t}] \approx\sum_{n_{\mathrm{act}}=\pm 1}p_{n_{\mathrm{act}}}\Braket{\rho_{\mathrm{nuc}}(q,p)\mathcal{P}_{\rho_{\mathrm{el}}}(\bm{S})B^{\mathrm{SH}}(\bm{S}(t),n_{\mathrm{act}}(t))}_{\mathrm{FL}} ,
\end{equation}
where $\rho_{\mathrm{nuc}}(q,p)$ is the Wigner distribution of the initial nuclear density [Eq.~\ref{eq:nuc_dens}] and $\mathcal{P}_{\rho_{\mathrm{el}}}(\bm{S})$ is a projector onto a single point of the Bloch sphere that corresponds to the initial electronic density matrix, $\hat{\rho}_{\mathrm{el}}(q)$ 
\begin{equation}
\label{eq:electronic_project}
\mathcal{P}_{\rho_{\mathrm{el}}}(\bm{S})=\delta\!\left(S_{x}-\sqrt{1-\Delta P^{2}}\right)\delta\!\left(S_{y}\right)\delta\!\left(S_{z}+\Delta P\right) .
\end{equation}
This Bloch sphere point is shown schematically in the bottom-left panel of Fig.~2 of the main paper. Additionally, the initial active surface for each trajectory is determined stochastically from the initial electronic wavefunction probabilities, $p_{\pm}=\tfrac{1}{2}(1\mp\Delta P)$.

In order to describe nonadiabatic transitions, the active surface is changed stochastically along the trajectory with a `hopping probability' given by the rate of change of the underlying electronic wavefunction
\begin{equation}
	P_{\pm\rightarrow\mp}=\pm\frac{2d(q)p}{m}\frac{Sx\delta t}{1\pm S_{z}} ,
\end{equation}
where negative probabilities are set to zero and the time-step, $\delta t$, is chosen to be small enough so that the probabilities are never greater than one. In the case of a successful hop, the nuclear momentum is rescaled along the nonadiabatic coupling vector (NACV) in order to conserve energy and reflected in the case of a frustrated hop.\footnote{For this model there is only a single nuclear degree of freedom and hence only one possible direction for applying the momentum rescaling.}

\subsubsection{The Mapping Approach to Surface Hopping (MASH)}
The mapping approach to surface hopping (MASH)\cite{Mannouch2023MASH} combines the best of both worlds of FSSH and mapping-based approaches. Unlike other mapping-based approaches, MASH has a surface-hopping force akin to FSSH. The active surface in MASH is not an additional parameter within the theory, but it is determined uniquely from the spin-mapping variables. For two-state systems 
\begin{equation}
	n_{\mathrm{act}}=\mathrm{sgn}(S_{z}) .
\end{equation}
This means that MASH has deterministic dynamics for which the electronic wavefunction and the active surface remain consistent throughout the dynamics. `Hops' therefore occur in MASH every time the spin vector evolves from one hemisphere to the other. One additional advantage of MASH over FSSH is that the underlying theory uniquely prescribes how the momentum rescaling should be applied at a hop, which corresponds to rescaling along the NACV and reflecting in the case of a frustrated hop. 

In order for the `hopping times' to differ for each trajectory, the initial spin-mapping variables must be sampled. To describe an initial electronic coherence, the spin-mapping variables are sampled over the entire Bloch sphere, weighted according to
\begin{equation}
	\Tr[\hat{\rho}\eu{i\hat{H}t}\hat{B}\eu{-i\hat{H}t}]\approx\tfrac{1}{2}\Braket{\rho_{\mathrm{nuc}}(q,p)\left[\mathcal{W}_{\mathrm{P}B}(\bm{S})\left(1 - \Delta P\,\mathrm{sgn}(S_{z})\right)+\mathcal{W}_{\mathrm{C}B}(\bm{S})\sqrt{1-\Delta P^{2}}\,S_{x}\right]B^{\mathrm{SH}}(\bm{S}(t))}_{\mathrm{FL}} ,
\end{equation}
The weighting factors, $\mathcal{W}_{AB}(\bm{S})$, are determined so that MASH both reproduces the short-time limit of exact quantum dynamics and the Landau-Zener transition rate in the fast nuclei limit
\begin{equation}
	\mathcal{W}_{AB}(\bm{S})=\begin{cases}
		3\qquad\quad\,\,\,\mathrm{when}\, A=B=\mathrm{C}, \\
		2\qquad\quad\,\,\,\mathrm{when}\, A=\mathrm{P},\,B=\mathrm{C}\,\,\mathrm{or}\,A=\mathrm{C},\,B=\mathrm{P} \\
		2|S_{z}|\qquad\mathrm{when}\, A=B=\mathrm{P}. \\
	\end{cases}
\end{equation}
where C refers to a coherence operator (i.e., $\hat{\sigma}_{x}$ and $\hat{\sigma}_{y}$) and P to a population operator (i.e., $\hat{\mathcal{I}}$, $\hat{\sigma}_{z}(q)$, or the adiabatic population operators themselves). 

\subsection{Mean-Field Approaches}
For mean-field approaches, the nuclei are propagated according to the mean-field force, which corresponds to the expectation value of the nuclear force operator with respect to the electronic wavefunction
\begin{equation}
\label{eq:force_MF}
\mathcal{F}^{\mathrm{MF}}(q,\bm{S})=-\frac{\partial \bar{V}(q)}{\partial q}-r_{\mathrm{s}}\left[\frac{\partial V_{z}(q)}{\partial q}S_{z}-2V_{z}(q)d(q)S_{x}\right] ,
\end{equation}
where $r_{\mathrm{s}}$ is the radius of the spin sphere.

The mean-field representation of the Pauli spin matrices for constructing electronic observables are given in terms of the electronic spin-mapping variables as
\begin{equation}
\sigma^{\mathrm{MF}}_{j}(\bm{S})=r_{\mathrm{s}}S_{j} . 
\end{equation}
In addition, for Ehrenfest dynamics and spin-LSC, $\mathcal{I}(\bm{S})=1$.
\subsubsection{Ehrenfest}
In Ehrenfest dynamics,\cite{Tully1990hopping} the electronic spin-mapping variables are initialized on the same point on the Bloch sphere ($r_{\mathrm{s}}=1$) as in FSSH
\begin{equation}
\label{eq:corr_ehren}
\Tr[\hat{\rho}\eu{i\hat{H}t}\hat{B}\eu{-i\hat{H}t}]\approx\Braket{\rho_{\mathrm{nuc}}(q,p)\mathcal{P}_{\rho_{\mathrm{el}}}(\bm{S})B^{\mathrm{MF}}(\bm{S}(t))}_{\mathrm{FL}} ,
\end{equation}
where $\mathcal{P}_{\rho_{\mathrm{el}}}(\bm{S})$ is given by Eq.~(\ref{eq:electronic_project}).
\subsubsection{Spin-LSC}
In spin-LSC,\cite{spinmap,multispin} the improved accuracy over Ehrenfest dynamics arises from introducing an initial sampling over the spin sphere, even when the dynamics correspond to starting in a pure electronic state. This is achieved by using a larger spin sphere ($r_{\mathrm{s}}=\sqrt{3}$), so that
\begin{equation}
\Tr[\hat{\rho}\eu{i\hat{H}t}\hat{B}\eu{-i\hat{H}t}]\approx\Braket{\rho_{\mathrm{nuc}}(q,p)\mathcal{P}_{\mathrm{LSC}}(\bm{S})\left[\left(1 - \sqrt{3}\Delta P S_{z}\right)+\sqrt{3}\sqrt{1-\Delta P^{2}}\,S_{x}\right]B^{\mathrm{MF}}(\bm{S}(t))}_{\mathrm{FL}} .
\end{equation}
The spin-LSC representation of the correlation function has the same structure as that for Ehrenfest dynamics [Eq.~\ref{eq:corr_ehren}], because $\tfrac{1}{2}\mathcal{P}_{\rho_{\mathrm{el}}}(\bm{S})\left[1-\Delta P S_{z}+\sqrt{1-\Delta P^{2}}S_{x}\right]=\mathcal{P}_{\rho_{\mathrm{el}}}(\bm{S})$.

Within spin-LSC, there are different options for the initial sampling of the spin-mapping variables. We choose to use so-called focused initial conditions, which sample the spin-mapping variables from the polar circles associated with the upper and lower hemispheres 
\begin{equation}
\label{eq:LSC_project}
\mathcal{P}_{\mathrm{LSC}}(\bm{S})=\frac{1}{2}\left[\delta\!\left(S_{z}-\frac{1}{\sqrt{3}}\right)+\delta\!\left(S_{z}+\frac{1}{\sqrt{3}}\right)\right] .
\end{equation}
These polar circles are shown schematically by the upper-middle panel of Fig.~2 of the main paper. By using such focused initial conditions, spin-LSC is guaranteed to reproduce the correct classical nuclear dynamics on a single surface in the Born-Oppenheimer limit and also incorporates the correct description of the quantum uncertainty between the spin degrees of freedom. 
\subsection{Spin-PLDM}
All of the methods introduced so far are so-called fully-linearized approaches,\cite{Miller2001SCIVR,linearized} which use a single set of spin-mapping variables to represent the single-time correlation function. Another class of quasiclassical methods, referred to as partially-linearized approaches,\cite{Huo2011densitymatrix,Huo2012PLDM,Hsieh2012FBTS,Hsieh2013FBTS} use separate spin-mapping variables to represent the forward and backward time propagators within the correlation function. The partially linearized version of spin-LSC is spin-PLDM,\cite{spinPLDM1,spinPLDM2} which represents the correlation function as follows
\begin{equation}
\Tr[\hat{\rho}\eu{i\hat{H}t}\hat{B}\eu{-i\hat{H}t}]\approx\Braket{\rho_{\mathrm{nuc}}(q,p)\mathcal{P}_{\mathrm{LSC}}(\bm{S}^{(\mathrm{f})})\mathcal{P}_{\mathrm{LSC}}(\bm{S}^{(\mathrm{b})})B^{\mathrm{MF}}(\bm{S}^{\mathrm{obs}}(t))}_{\mathrm{PL}} ,
\end{equation}
where $\bm{S}^{(\mathrm{f})}$ and $\bm{S}^{(\mathrm{b})}$ are the spin-mapping variables for the forward and backward propagation respectively. These mapping variables are sampled from the same polar circles as in spin-LSC (Eq.~\ref{eq:LSC_project}), which means that the centroid, $\bm{S}=\tfrac{1}{2}(\bm{S}^{(\mathrm{f})}+\bm{S}^{(\mathrm{b})})$ is distributed according to the upper-right panel of Fig.~2 of the main paper. In spin-PLDM, the centroid is the dynamical variable that determines the nuclear force (via Eqs.~(\ref{eq:dynamics}) and (\ref{eq:force_MF}) with $r_{\mathrm{s}}=\sqrt{3}$).

In addition, the observables in spin-PLDM are determined by another set of spin-mapping variables, $\bm{S}^{\mathrm{obs}}$, related to $\bm{S}^{(\mathrm{f})}$ and $\bm{S}^{(\mathrm{b})}$ by  
\begin{equation}
\label{eq:spin_obs}
S_{j}^{\mathrm{obs}}(t)=\frac{4}{\sqrt{3}}\mathrm{Re}\!\left[\tr\!\left[\hat{\rho}_{\mathrm{el}}(q)\hat{w}\!\left(\bm{S}^{(\mathrm{b})}\right)\hat{U}^{\dagger}(t)\hat{\sigma}_{j}(q)\hat{U}(t)\hat{w}\!\left(\bm{S}^{(\mathrm{f})}\right)\right]\right] .
\end{equation}
Here $\tr[\cdots]$ refers to the quantum trace over the electronic degrees of freedom, $\hat{U}$ is the time-ordered electronic propagator along the nuclear path and $\hat{w}(\bm{S})$ is the Stratonovich-Weyl kernel given by
\begin{equation}
\hat{w}(\bm{S})=\frac{1}{2}\left[\hat{\mathcal{I}}+\sqrt{3}\bm{S}\cdot\hat{\bm{\sigma}}(q)\right] .
\end{equation}

While Eq.~(\ref{eq:spin_obs}) is in principle sufficient for calculating $\bm{S}^{\mathrm{obs}}(t)$, we propose an easier scheme in practice for propagating these spin-mapping variables along the trajectory. Firstly, from Eq.~(\ref{eq:spin_obs}) it can be shown that $\bm{S}^{\mathrm{obs}}$ obeys the analogous equations of motion to Eqs.~(\ref{eq:dynam_Sx}), (\ref{eq:dynam_Sy}) and (\ref{eq:dynam_Sz}). Namely
\begin{subequations}
\label{eq:dynamics_obs}
\begin{align}
\dot{S}^{\mathrm{obs}}_{x}&=\frac{2d(q)p}{m}S^{\mathrm{obs}}_{z}-2V_{z}(q)S^{\mathrm{obs}}_{y} , \\
\dot{S}^{\mathrm{obs}}_{y}&=2V_{z}(q)S^{\mathrm{obs}}_{x} ,  \\
\dot{S}^{\mathrm{obs}}_{z}&=-\frac{2d(q)p}{m}S^{\mathrm{obs}}_{x} . 
\end{align}
\end{subequations}
Given that their $t=0$ values can be determined from Eq.~(\ref{eq:spin_obs}) as follows
\begin{subequations}
\label{eq:obs_t=0}
	\begin{align}
        \begin{split}
		S_{x}^{\mathrm{obs}}(0)&=\left(S^{(\mathrm{f})}_{x}+S^{(\mathrm{b})}_{x}\right)-\sqrt{3}\Delta P\left(S^{(\mathrm{f})}_{z}S^{(\mathrm{b})}_{x}+S^{(\mathrm{b})}_{z}S^{(\mathrm{f})}_{x}\right)+\sqrt{3}\sqrt{1-\Delta P^{2}}\left(2S^{(\mathrm{f})}_{x}S^{(\mathrm{b})}_{x} + \frac{1}{3}-\bm{S}^{(\mathrm{f})}\cdot\bm{S}^{(\mathrm{b})}\right) 
        \end{split}, \\
		S_{y}^{\mathrm{obs}}(0)&=\left(S^{(\mathrm{f})}_{y}+S^{(\mathrm{b})}_{y}\right)-\sqrt{3}\Delta P\left(S^{(\mathrm{f})}_{z}S^{(\mathrm{b})}_{y}+S^{(\mathrm{b})}_{z}S^{(\mathrm{f})}_{y}\right)+\sqrt{3}\sqrt{1-\Delta P^{2}}\left(S^{(\mathrm{f})}_{x}S^{(\mathrm{b})}_{y} + S^{(\mathrm{b})}_{x}S^{(\mathrm{f})}_{y}\right) , \\
		\begin{split}
		S_{z}^{\mathrm{obs}}(0)&=\left(S^{(\mathrm{f})}_{z}+S^{(\mathrm{b})}_{z}\right)-\sqrt{3}\Delta P\left(2S^{(\mathrm{f})}_{z}S^{(\mathrm{b})}_{z}+\frac{1}{3}-\bm{S}^{(\mathrm{f})}\cdot\bm{S}^{(\mathrm{b})}\right)+\sqrt{3}\sqrt{1-\Delta P^{2}}\left(S^{(\mathrm{f})}_{x}S^{(\mathrm{b})}_{z}+S^{(\mathrm{b})}_{x}S^{(\mathrm{f})}_{z}\right) 
        \end{split},
	\end{align}
\end{subequations}
Eqs.~(\ref{eq:dynamics_obs}) and (\ref{eq:obs_t=0}) therefore provide a simpler way of calculating $\bm{S}^{\mathrm{obs}}(t)$ without the need to explicitly evaluate Eq.~(\ref{eq:spin_obs}).

Finally, noting that $\hat{U}^{\dagger}(t)\hat{U}(t)=\hat{\mathcal{I}}$, the spin-PLDM representation of the identity operator is given by  
\begin{equation}
\begin{split}
\mathcal{I}^{\mathrm{PLDM}}&=4\mathrm{Re}\!\left[\tr\!\left[\hat{\rho}_{\mathrm{el}}(q)\hat{w}\!\left(\bm{S}^{(\mathrm{b})}\right)\hat{w}\!\left(\bm{S}^{(\mathrm{f})}\right)\right]\right]  \\
&=\left(1 + 3\bm{S}^{(\mathrm{f})}\cdot\bm{S}^{(\mathrm{b})}\right)-\sqrt{3}\Delta P\left(S^{(\mathrm{f})}_{z}+S^{(\mathrm{b})}_{z}\right)+\sqrt{3}\sqrt{1-\Delta P^{2}}\left(S^{(\mathrm{f})}_{x}+S^{(\mathrm{b})}_{x}\right) . 
\end{split}
\end{equation}
\section{Evaluation of the Coherence Measures}
In order to evaluate the coherence measures introduced in the main paper with quasiclassical trajectory approaches, the trajectories are first used to compute the correlation functions associated with $\hat{B}=\hat{\sigma}_{x}(q)$, $\hat{\sigma}_{y}(q)$, $\hat{\sigma}_{x}(q)\ket{q}\!\bra{q}$ and $\hat{\sigma}_{y}(q)\ket{q}\!\bra{q}$. The coherence measures are then constructed in terms of the correlation functions using the formulas provided in the main paper.

In order to calculate the correlation functions involving the nuclear projection operator, $\hat{P}_{q}=\ket{q}\!\bra{q}$, we note that the Wigner transform of this quantity is
\begin{equation}
P^{\mathrm{W}}_{q}(q'(t),p'(t))=\delta(q-q'(t)) .
\end{equation}
This can be evaluated by histogramming the trajectories.

For the numerical simulations, $10^{6}$ trajectories were used with a timestep of $0.25$~fs. All trajectories of the mapping-based approaches were repeated once with the initial conditions switched as follows: $S_{x}\rightarrow-S_{x}$ and $S_{y}\rightarrow-S_{y}$. For spin-PLDM, this replacement was performed independently for $\bm{S}^{(\mathrm{b})}$ and $\bm{S}^{(\mathrm{f})}$, meaning that the trajectories were actually repeated four times in this case. The $\mathcal{C}_{-+}(t)$ coherence measure was computed using equally spaced bins of width $\Delta q=0.005$ between $q=-7$ and $q=7$. A large number of trajectories was used to make sure that the $\mathcal{C}_{-+}(t)$ coherence measure was sufficiently converged for such a fine grid. Far fewer trajectories would have been needed to compute the other observables, as well as this coherence measure on a coarser grid.

\newpage

\clearpage
\bibliographystyle{achemso}
\bibliography{references,extra_refs} 